\documentclass[12pt]{iopart}
\usepackage{lscape}
\usepackage{iopams}
\usepackage{graphicx}
\usepackage{ulem}
\usepackage[ps2pdf,hypertex]{hyperref}
\usepackage{datetime}

\DeclareMathAlphabet{\mathpzc}{OT1}{pzc}{m}{it}

\begin{document}
\title{Tensorial form and matrix elements of the relativistic nuclear recoil operator}
\author{E. Gaidamauskas$^{1}$,  C. Naz\'e$^{2}$, P. Rynkun$^{3}$, G. Gaigalas$^{1,3}$, P.~J\"onsson$^{4}$  and M. Godefroid$^{2}$}

\address{$^{1}$ Vilnius University Research Institute of Theoretical Physics and Astronomy, \\
A. Go\v{s}tauto 12, LT-01108 Vilnius, Lithuania}

\address{$^{2}$ Chimie Quantique et Photophysique, CP160/09, Universit\'e Libre de Bruxelles, \\
Av. F.D. Roosevelt 50, B-1050 Brussels, Belgium}

\address{$^{3}$ Vilnius Pedagogical University, \\
Student\c u 39, LT-08106 Vilnius, Lithuania}

\address{$^{4}$ School of Technology, Malm\"{o} University, 205-06 Malm\"{o}, Sweden }

\eads{erikas.gaidamauskas@tfai.vu.lt, cnaze@ulb.ac.be}
\begin{abstract}
Within the lowest-order relativistic approximation ($\sim v^2/c^2$) and to first order in $m_e/M$, the tensorial form of the relativistic  corrections of the nuclear recoil Hamiltonian is derived, opening interesting perspectives for calculating isotope shifts in the multiconfiguration Dirac-Hartree-Fock framework.  Their calculation is illustrated for selected Li-, B- and C-like ions. The present work underlines the fact that the relativistic corrections to the nuclear recoil are definitively necessary for getting reliable isotope shift values.  

\end{abstract}

\submitto{\jpb}

\vfill

\noindent \hfill  \today @ \currenttime

\maketitle

\section{Introduction}


Nuclear and relativistic effects in atomic spectra are treated in the pioneer works of Stone~\cite{Sto:61a,Sto:63a} and~Veseth \cite{Ves:85a}. The theory of the mass shift has then been reformulated by Palmer~\cite{Pal:87a}. Calculations of nuclear motional effects in many-electron atoms have been performed by Parpia and co-workers~\cite{Paretal:92a,JonFro:97a} in the relativistic scheme, using fully relativistic wave functions, but adopting the non-relativistic form of the recoil operator. Relativistic nuclear recoil corrections to the energy levels of multicharged ions have been estimated by Shabaev and Artemyev~\cite{ShaArt:94a} who derived the relativistic corrections of the recoil Hamiltonian. In a study of isotope shifts of forbidden transitions in Be- and B-like argon ions, 
Tupitsyn \etal~\cite{Tupetal:03a} showed that a proper evaluation of the mass isotope shift requires the use of this relativistic recoil operator.  The latter  has also been shown to be crucial by Porsev \etal~\cite{Poretal:09a} for calculating isotope shifts of transitions between the fine structure energy levels of the ground multiplets of Fe~I and Fe~II. 

As far as computational atomic structure is concerned, the extension of the available relativistic codes such as \textsc{grasp2k}~\cite{Jonetal:07a} or \textsc{mcdf}-gme~\cite{IndDes:08a,Des:93a} is needed for estimating these mass corrections properly for any many-electron system. Programs to calculate pure angular momentum coefficients for any scalar one- and two- particle operator are  available~\cite{Gaietal:01a} but do require the knowledge of the tensorial structure of the operators to be integrated between the many-electron atomic wave functions \cite{Gra:07a}.
The tensorial form of the nuclear recoil Hamiltonian is derived in the present work, opening interesting perspectives for calculating isotope shifts in the multiconfiguration Dirac-Hartree-Fock 
~(MCDHF) framework.


\section{The relativistic mass shift operator}

In the MCDHF method, the atomic state function (ASF)
$\Psi (\gamma P J M_J)$, of a stationary state of an atom, is expressed as a linear combination of
symmetry-adapted configuration state functions (CSFs) $\Phi (\gamma_{p} P J M_J)$,~i.e.
\begin{eqnarray}
\label{eq:asf}
\Psi (\gamma P J M_J) = \sum_{p} c_{p} \Phi (\gamma_{p} P J M_J),
\end{eqnarray}
where $J$ is the total electronic angular momentum of the state,  $\gamma$ represents the electronic configuration and intermediate quantum numbers, and $P$ stands for the parity. The mixing coefficients $c_p$ and the one-electron radial wave functions spanning the CSFs are optimized by solving the MCDHF equations iteratively until self-consistency. The latter are derived by applying the variational principle to the  energy functional based on the Dirac-Coulomb Hamiltonian \cite{Gra:07a}
\begin{eqnarray}
\label{eg:MCDF}
H_{DC}  =
\sum_{i=1}^N
\Big( c \boldsymbol \alpha_i \cdot \boldsymbol  p_i + (\beta_i-1)c^2 +V({r_i})\Big)
 + \sum_{i < j}^N
\frac{1}{r_{ij}},
\end{eqnarray}
where $V({r_i})$ is the monopole part of the electron-nucleus interaction, $\boldsymbol \alpha$ and $\beta$ are the $(4\times 4)$ Dirac matrices and $c$ is the speed of light ($c = 1/\alpha $ in atomic units, where $\alpha$ is the fine-structure constant).

The mass shift of the energy levels in an atom with nuclear mass $M$ is caused by the recoil motion of the atomic nucleus. The corresponding recoil Hamiltonian 
\begin{eqnarray}
\label{eq:H_MS}
   {H}_{\rm MS}  \; = \;
 \frac{1}{2M}
 \sum_{i,j} ^{N} \;
\left( \boldsymbol { p}_{i} \cdot \boldsymbol {p}_{j} - \frac{\alpha Z}{r_i} \left(\boldsymbol \alpha_{i} + \frac{\left(\boldsymbol \alpha_i\cdot\boldsymbol  r_i\right) \boldsymbol r_i}{r^{2}_{i}}\right)\cdot \boldsymbol  p_j\right) \; ,
\end{eqnarray}
has been derived within the lowest-order relativistic approximation and to first order  in $m/M$ by Shabaev and collaborators \cite{ShaArt:94a,Tupetal:03a}.
Rewriting  it as the sum of the normal mass shift (NMS) and specific mass shift (SMS) contributions 
and using the tensorial form $\boldsymbol r^{1} = r\boldsymbol C^{1}$,~\eref{eq:H_MS} becomes
\begin{eqnarray}
\label{eq:separation}
{H}_{\rm MS} = {H}_{\rm NMS} + {H}_{\rm SMS} \; ,
\end{eqnarray}
with
\begin{eqnarray}
\label{eq:H_NMS}
   {H}_{\rm NMS}  \; = \;
 \frac{1}{2M}
 \sum_{i=1} ^{N} \;
\left( \boldsymbol { p}^{2}_{i} 
- \frac{\alpha Z}{r_i} \boldsymbol \alpha_{i} \cdot\boldsymbol  p_i 
- \frac{\alpha Z}{r_i} \left(\boldsymbol \alpha_i\cdot\boldsymbol  C^{1}_{i}\right) \boldsymbol C^{1}_{i} \cdot\boldsymbol  p_i
\right), \;
\end{eqnarray}
\begin{eqnarray}
\label{eq:H_SMS}
 {H}_{\rm SMS}  \; &= &\;
 \frac{1}{M}
 \sum_{i < j} ^{N} \;
\left( 
\boldsymbol { p}_{i} \cdot \boldsymbol {p}_{j} 
- \frac{\alpha Z}{r_i} \boldsymbol \alpha_{i} \cdot\boldsymbol  p_j 
- \frac{\alpha Z}{r_i} \left(\boldsymbol \alpha_i\cdot\boldsymbol  C^{1}_{i}\right) \boldsymbol C^{1}_{i} \cdot\boldsymbol  p_j\right) \; ,
\end{eqnarray}
that, in both cases, are rewritten as a sum of three separate contributions:
\begin{eqnarray} 
\label{eq:H_NMS_split}
&{H}_{\rm NMS}&\equiv H^1_{\rm NMS} + H^2_{\rm NMS} + H^3_{\rm NMS}\;,
\end{eqnarray}
and
\begin{eqnarray}
\label{eq:H_SMS_split}
  &{H}_{\rm SMS}&\equiv H^1_{\rm SMS} + H^2_{\rm SMS} + H^3_{\rm SMS} 
\; .
\end{eqnarray}
Since the expectation values of the NMS and SMS operators are evaluated with the MCDHF wave functions, the expectation values $\langle H^1_{\rm NMS}\rangle$ and $\langle H^1_{\rm SMS}\rangle$ partly contain the relativistic contributions. Tupitsyn \etal \cite{Tupetal:03a} pointed out that averaging the non-relativistic recoil operator with the relativistic wave functions strongly overestimates the relativistic correction to the recoil effect such as it becomes important to use the complete form \eref{eq:H_MS} when one works in the relativistic scheme.  
\subsection{Normal mass shift expectation value}
The (mass-independent) normal mass shift parameter $K_{\rm NMS}$ is defined by the following expression
\begin{eqnarray}
\label{eq:NMS}
   \frac{K_{\rm NMS}}{M} \; \equiv    \langle \Psi (\gamma P J M_J)|
   {H}_{\rm NMS}| \Psi (\gamma P J M_J)\rangle \; .
\end{eqnarray}
By analogy with \eref{eq:H_NMS_split}, we  define $K_{\rm NMS}$ as the sum of $K_{\rm NMS}=K^1_{\rm NMS} + K^2_{\rm NMS} + K^3_{\rm NMS}$.
Applying the Wigner-Eckart theorem \cite{Cow:81a}, the matrix element of the normal mass shift operator is $M_J$-invariant and is proportional to the reduced matrix element (r.m.e.)\footnote{The two definitions of r.m.e are related to each other through 
$ \langle \gamma J' \| O^k \| \gamma  J \rangle  =  \sqrt{2J'+1} \; [ \gamma J' \| O \| \gamma  J ] $.}
\begin{eqnarray}
\label{eq:NMSred}
   \frac{K_{\rm NMS}}{M} \; & = &   
   \frac{1}{\sqrt{2J+1}} \; \langle \Psi (\gamma P J) \| {H}_{\rm NMS}\| \Psi (\gamma P J)\rangle   
 \nonumber \\
    & = &   
   \left[ \Psi (\gamma P J)\|
  {H}_{\rm NMS}\| \Psi (\gamma P J)\right]
 \; .
\end{eqnarray}
Using multiconfiguration expansions \eref{eq:asf}, the reduced matrix elements
of the general spherical tensor operator ${T}^{k}_{q}$ becomes
\begin{eqnarray}
\label{eq:mat_element_between_ASF}
   \left[ \Psi (\gamma P J) \| \boldsymbol {T}^{k} \| \Psi (\gamma P J) \right]
   \; = \; \sum_{p,s} \; c_p c_s
   \left[ \Phi  ( \gamma_p P J ) \| \boldsymbol {T}^{k} \| \Phi ( \gamma_s P J ) \right].
\end{eqnarray}
The reduced matrix elements of the one-electron operator  
$\boldsymbol {T}^{k} = \sum_i \boldsymbol {t}^{k} (i)$  between CSFs is expressed as a sum over single-particle reduced matrix elements
\begin{eqnarray}
\label{eq:mat_element_between_CSF}
   \left[ \Phi ( \gamma_p P J ) \| \boldsymbol {T}^{k} \| \Phi ( \gamma_s P J) \right] =
   \; \sum_{a,b} \; T_{ps}(ab)  \left[ n_a \kappa_a \| \boldsymbol {t}^{k} \| n_b \kappa_b \right] \; ,
\end{eqnarray}
where the $T_{ps}(ab)$ are the spin-angular coefficients arising from Racah's algebra~\cite{Gaietal:01a,Gra:07a,Gaietal:97a}.
Introducing the one-body normal mass shift operator associated to~\eref{eq:H_NMS} ($H_{\rm NMS}=\sum_i{h}_{\rm NMS}(i)$)
\begin{eqnarray}
\label{eq:h_NMS}
   {h}_{\rm NMS}  \; = \;
 \frac{1}{2M}
  \;
\left( \boldsymbol { p}^{2} - \frac{\alpha Z}{r} \left(\boldsymbol \alpha + \left(\boldsymbol \alpha\cdot\boldsymbol  C^{1}\right) \boldsymbol C^{1}\right)\cdot\boldsymbol  p\right) \; ,
\end{eqnarray}
we hereafter derive the expression of  its r.m.e.,  using relativistic central-field one-electron wave functions
\begin{eqnarray}
\label{eq:oew}
\psi_{n_a, \kappa_a, m_a} (\boldsymbol r, \sigma) = \frac{1}{r} \;
\left(
   \begin{array}{c}
      P_{n_a,\kappa_a}(r)\Omega_{\kappa_a, m_a}(\theta, \phi, \sigma)  \\
     i Q_{n_a,\kappa_a}(r)\Omega_{-\kappa_a, m_a}(\theta, \phi, \sigma) 
   \end{array}
   \right).
\end{eqnarray}
 $P_a$ and $Q_a$ are respectively  the large and small components of the relativistic one-electron radial wave function $a=(n_a \kappa_a)$, where $\kappa = (l-j)(2j+1)$. \\

Introducing the notation $\partial_r \equiv \frac{\partial}{\partial r}$, the action of the operator $\boldsymbol p^2$ on the  large $(F=P)$ and the small $(F=Q)$ component of a relativistic wave function 
\begin{eqnarray}
\label{eq:p1}
\boldsymbol p^2 \frac{F_{n, \kappa}(r)}{r}\Omega_{\kappa, m} (\theta, \phi, \sigma) = \frac{1}{r}\left(-\partial^{2}_{r}+\frac{l(l+1)}{r^2}\right)F_{n, \kappa}(r)\Omega_{\kappa, m} (\theta, \phi, \sigma) 
\end{eqnarray}
is found using 
\begin{eqnarray}
\label{eq:p}
\boldsymbol p^2 =-\Delta = - \frac{1}{r^2} \partial_r r^2 \partial_r + \frac{\bi{l} ^2}{r^2} \; .
\end{eqnarray}
From this expression and integrating by parts, the first term of the one-electron matrix element NMS operator \eref{eq:h_NMS} becomes,
\begin{eqnarray}
\label{eq:mat_NMS1}
   \langle n_a \kappa_a m_a |
  \frac{\boldsymbol p^2}{2} | n_b \kappa_b m_b\rangle
 = 
\delta(\kappa_a m_a, \kappa_b m_b) 
 \\
  \times \frac{1} {2}\; 
   \int_{0}^{\infty} dr
   \Bigg( \left(\partial_{r} P_a\right) \left(\partial _{r}  P_b\right) \; + \; \left(\partial_{r} Q_a\right) \left(\partial_{r}  Q_b \right)
    +  \frac{l_b(l_b+1) P_a P_b + \widetilde{l}_{b}(\widetilde{l}_{b} + 1) Q_a Q_b}{r^2}\Bigg),  \nonumber
\end{eqnarray}
with $\widetilde{l} = 2j - l$.
Building the Dirac matrices $\boldsymbol \alpha $ from $\boldsymbol \alpha = \sigma_x \otimes \boldsymbol \sigma$  with
\begin{eqnarray}
\label{eq:s}
   \sigma_x = 
\left( 
\begin{array}{cc}
      0 & 1  \\
     1 & 0
   \end{array}
   \right),
\end{eqnarray}
one rewrites the second and the third parts of the NMS operator \eref{eq:h_NMS} as
\begin{eqnarray}
\label{eq:A}
\left(\frac{-\alpha Z}{2r}\right)\left(\boldsymbol \alpha\cdot \boldsymbol p + \left(\boldsymbol \alpha\cdot \boldsymbol  C^{1}\right) \left(\boldsymbol C^{1} \cdot\boldsymbol  p\right)\right) \equiv
 \sigma_x \otimes A = \left( 
\begin{array}{cc}
      0 & A  \\
     A &  0
   \end{array}
   \right)
\end{eqnarray}
with
\begin{eqnarray}
\label{eq:A1}
 A = \left(\frac{-\alpha Z}{2r}\right)\left(\boldsymbol \sigma \cdot\boldsymbol p + \left(\boldsymbol \sigma \cdot\boldsymbol  C^{1} \right) \left(\boldsymbol C^{1}\cdot \boldsymbol  p\right)\right).
\end{eqnarray}
Taking into account that (see (A.4.9) and (3.2.14) in \cite{Gra:07a})
\begin{eqnarray}
\label{eq:r}
 \boldsymbol \sigma \cdot\boldsymbol  C^{1} =  \boldsymbol \sigma \cdot\boldsymbol  e_r = \sigma_r
 \hspace*{0.5 cm} ; \hspace*{0.5 cm}  \boldsymbol C^{1} \cdot\boldsymbol  p =\boldsymbol e_r \cdot\boldsymbol  p = (-i\partial_r),
\end{eqnarray}
and
\begin{eqnarray}
\label{eq:r1}
\boldsymbol \sigma\cdot \boldsymbol p = -i\sigma_r \left(\partial_r + \frac{K+1}{r}\right),
\end{eqnarray}
with $K = - (1 +  \boldsymbol \sigma\cdot \bi{l} ) $,
the operator A becomes
\begin{eqnarray}
\label{eq:A2}
 A = \left(\frac{-\alpha Z}{2r}\right)\left(-i\sigma _r\right)\left( 2\partial_r + \frac{K+1}{r}\right).
\end{eqnarray}
Acting on the one-electron relativistic wave function component, it gives
\begin{eqnarray}
\label{eq:p1_2}
A \frac{F_{n, \kappa}(r)}{r}\Omega_{\kappa, m} = \left(\frac{-\alpha Z}{2r}\right)\frac{i}{r}\left(2\partial_{r}+\frac{\kappa-1}{r}\right)F_{n, \kappa}(r)\Omega_{-\kappa, m} \; ,
\end{eqnarray}
from which one derives, integrating by parts, the one-electron matrix element of the second and third parts of the NMS operator
\begin{eqnarray}
\label{eq:mat_NMS2}
\langle n_a \kappa_a m_a |
  \sigma_x \otimes A | n_b \kappa_b m_b\rangle
=
\delta(\kappa_a m_a, \kappa_b m_b)
  \\
 \times \frac{1} {2} \int_{0}^{\infty}
\Bigg( (-2 \alpha Z)\frac{Q_a \partial_r P_b + Q_b \partial_r P_a}{r}+(- \alpha Z)\left(\frac{\kappa_b-1}{r^2}\right)(P_b Q_a+ P_a Q_b)\Bigg)  dr \nonumber
\end{eqnarray}
Combining \eref{eq:mat_NMS1}  and   \eref{eq:mat_NMS2} to deduce the r.m.e. of the normal mass shift operator~\eref{eq:h_NMS}, we obtain the final expression 
\begin{eqnarray}
\label{eq:mat_NMS} 
& & \left[ n_a \kappa_a \|
{h}_{\rm NMS} \| n_b \kappa_b \right]  =  
\delta( \kappa_a, \kappa_b )  \\
& \times & \frac{1} {2M}
\int_{0}^{\infty}\Bigg( \left(\partial_{r} P_a\right) \left(\partial _{r}  P_b\right) + \left(\partial_{r} Q_a\right) \left(\partial_{r}  Q_b\right) +  \frac{l_b(l_b+1) P_a P_b + \widetilde{l}_{b}(\widetilde{l}_{b} + 1) Q_a Q_b}{r^2}   \nonumber \\ 
& + & \left(-2\alpha Z\right) \frac{Q_a \partial_r P_b + Q_b \partial_r P_a}{r}
+ (-\alpha Z)\left(\frac{\kappa_b-1}{r^2}\right)(Q_a P_b + Q_b P_a)\Bigg)dr.
   \nonumber 
\end{eqnarray}

\subsection{Specific mass shift expectation value}
Similarly to \eref{eq:NMSred}, the (mass-independent) specific mass shift parameter $K_{\rm SMS}$ is defined as
\begin{eqnarray}
\label{eq:SMSred}
   \frac{K_{\rm SMS}}{M} \equiv 
   \left[ \Psi (\gamma P J)\|
  {H}_{\rm SMS}\| \Psi (\gamma P J)\right] \; ,
\end{eqnarray}
and $K^1_{\rm SMS}, K^2_{\rm SMS}, K^3_{\rm SMS}$ as its contributions according to \eref{eq:H_SMS_split}.
Its evaluation requires the calculation of the corresponding matrix elements in the CSF space. For the
 general scalar two-particle operator  
\begin{eqnarray}\label{eq:G}
G = \sum_{i < j} g(i,j) 
\end{eqnarray}
with
\begin{eqnarray}\label{eq:g1}
g(i,j) = \sum_k g_k (r_i, r_j) \left(\mathbf T^k(i) \cdot \mathbf T^k(j)\right) \; ,
\end{eqnarray}
 the reduction of the many-electron r.m.e. 
 in terms of the two-electron integrals $X^{k}$, also called \textit{effective interaction strengths}~\cite{Gra:07a},
\begin{eqnarray}
\label{eq:mat_element_between_CSF_2}
   [ \Phi ( \gamma_p P J ) \| \sum_{i<j} g(i,j) \| \Phi ( \gamma_s P J)] =
   \;  \sum_{abcd} \; \sum_k   v^{(k)}_{ps}(abcd)  X^{k} (abcd) 
\end{eqnarray}
can be performed using Racah's algebra~\cite{Gaietal:01a,Gra:07a,Gaietal:97a}.  For the specific mass shift Hamiltonian~\eref{eq:H_SMS}, using $k=1$, all three terms  have the particular form
\begin{equation}
\label{eq:g_ij_gfact}
g(i,j) = g_1(r_i) g_1(r_j) \left(\mathbf T^1(i) \cdot \mathbf T^1(j)\right)
\end{equation}
in which the radial part $g_1(r_i,r_j)$ of \eref{eq:g1} is factorized.
Adopting the covariant notation for the $3j$-symbol of Wigner \cite{Wig:59a} and using the definition of the scalar product of two irreducible tensor operators  and the Wigner-Eckart theorem, the matrix element of \eref{eq:g_ij_gfact} can be written  as follows
\begin{eqnarray}
\label{eq:mat_SMS1}
   \langle a b |
  g (i, j) | c d\rangle 
   \; = \; \sum_{q = -1}^{1}
  \left( 
\begin{array}{ccc}
     m_a  & 1 & j_c \\
     j_a &  q &  m_c   
   \end{array}
   \right)
  \left( 
\begin{array}{ccc}
     m_b &  q &  j_d \\
     j_b &  1 &  m_d   
   \end{array}
   \right)
X^{1}(abcd),
\end{eqnarray}
where 
\begin{eqnarray}
\label{eq:mat_SMS2}
X^{1} (abcd)
    = -    \langle a \|
  g (r_i) \boldsymbol T^{1}(i) \| c \rangle  
 \langle b \|
   g (r_j) \boldsymbol T^{1}(j) \| d \rangle \; .
\end{eqnarray}
From the structure of~\eref{eq:H_SMS_split}, the latter has three components 
\begin{eqnarray}
\label{eq:mat_SMS2_2}
X^{1} (abcd)
   =  
X^{1}_{1} (abcd) + X^{1}_{2} (abcd) +  X^{1}_{3} (abcd) 
\end{eqnarray}
that we analyzed hereafter separately.
\subsubsection{First part: $X^{1}_{1} (abcd)$}
~\\

\noindent
Building $X^{1}_{1} (abcd)$ from \eref{eq:mat_SMS2},  we have
\begin{eqnarray}
\label{eq:mat_SMS3}
g (r_i) \boldsymbol T^{1}(i)   =  \boldsymbol p^{1} (i)
\hspace*{0.5cm} ; \hspace*{0.5cm}
g (r_j) \boldsymbol T^{1}(j)= \boldsymbol p^{1} (j) \; .
\end{eqnarray}
Introducing the one-electron reduced matrix element
\begin{eqnarray}
\label{eq:mat_SMS4}
\langle a \|
  \boldsymbol p^{1}  \| c \rangle  
   =  
 -i   
 \langle \kappa_ {a} \|
    \boldsymbol C^{1} \| \kappa_{c} \rangle \mathpzc{V} (n_{a} \kappa_{a} , n_{c} \kappa_{c} ) ,
\end{eqnarray}
where $\mathpzc{V} (n \kappa , n' \kappa')$ is the Vinti radial integral 
\begin{eqnarray}
\label{vinti}
\mathpzc{V} (n \kappa , n' \kappa')
&=&\int^\infty_0 P_{n\kappa}(r)\left[\frac{d}{dr}-\frac{\kappa(\kappa+1)-\kappa'(\kappa'+1)}{2r}\right]P_{n'\kappa'}(r)
dr \\
&+&\int^\infty_0 Q_{n\kappa}(r)\left[\frac{d}{dr}-\frac{-\kappa(-\kappa+1)+\kappa'(-\kappa'+1)}{2r}\right]Q_{n'\kappa'}(r)
dr
\nonumber \; ,
\end{eqnarray}
the first contribution to the effective interaction strength writes as
\begin{eqnarray}
\label{eq:mat_SMS5}
X^{1}_{1} (abcd)
   =    
 \langle \kappa_ {a} \|
    \boldsymbol C^{1} \| \kappa_{c} \rangle  \langle \kappa_ {b} \|
    \boldsymbol C^{1} \| \kappa_{d} \rangle \mathpzc{V} (n_{a} \kappa_{a} , n_{c} \kappa_{c} ) 
\mathpzc{V} (n_{b} \kappa_{b} , n_{d} \kappa_{d} ) \; ,
\end{eqnarray}
recovering the uncorrected relativistic expression used in \cite{Paretal:92a,JonFro:97a}.
\subsubsection{Second part: $X^{1}_{2} (abcd)$}~\\

\noindent For the second term  of the SMS operator, we identify from \eref{eq:H_SMS}, \eref{eq:H_SMS_split},
\eref{eq:mat_SMS2} and \eref{eq:mat_SMS2_2}
\begin{eqnarray}
\label{eq:mat_SMS6}
 g (r_i) \boldsymbol T^{1}(i)
   = \frac{- \alpha Z}{r} \left( \sigma_x \otimes \boldsymbol \sigma^{1} (i) \right) 
   \hspace*{0.5cm} ; \hspace*{0.5cm}
   g (r_j) \boldsymbol T^{1}(j) = \boldsymbol p^{1}(j) \; .
\end{eqnarray}
Introducing the matrix element 
\begin{eqnarray}
\label{eq:mat_SMS4_2}
\langle a |
r^{-1} \left( \sigma_x \otimes \boldsymbol \sigma^{1}_q  \right) 
  | c \rangle  \nonumber \\
 = i \int_{0}^{\infty}  \frac{dr}{r}  \left(-Q_a P_c \langle - \kappa_ {a} m_a |
    \boldsymbol \sigma^{1}_q | \kappa_{c} m_c \rangle 
 +   Q_c P_a \langle \kappa_ {a} m_a |
    \boldsymbol \sigma^{1}_q | - \kappa_{c} m_c \rangle \right),
\end{eqnarray}
and using the r.m.e.~\eref{eq:mat_SMS4}, the corresponding contribution to the effective interaction strength
is
\begin{eqnarray}
\label{eq:mat_SMS8}
&&X^{1}_{2} (abcd)=  -\langle \kappa_ {b} \|
    \boldsymbol C^{1} \| \kappa_{d} \rangle 
\mathpzc{V} (n_{b} \kappa_{b} , n_{d} \kappa_{d} )
 \nonumber\\
&  \times & \int_{0}^{\infty} dr
\left(\frac{- \alpha Z}{r}\right) \left( -Q_a P_c \langle - \kappa_ {a} \|
    \boldsymbol \sigma^{1} \| \kappa_{c} \rangle + Q_c P_a \langle \kappa_ {a} \|
    \boldsymbol \sigma^{1} \| - \kappa_{c} \rangle \right).
\end{eqnarray}

\subsubsection{Third part: $X^{1}_{3} (abcd)$}~\\

\noindent
Similarly, the two components of the third term of the SMS operator are
\begin{eqnarray}
\label{eq:mat_SMS6_2}
 g (r_i) \boldsymbol T^{1}(i)    = \frac{- \alpha Z}{r} \left(\sigma_x \otimes (\sigma_{r}(i) \boldsymbol C^{1} (i))\right)   \hspace*{0.5cm} ; \hspace*{0.5cm} 
 g (r_j) \boldsymbol T^{1}(j) =  \boldsymbol p^{1}(j) \; .
\end{eqnarray}
Using the matrix element 
\begin{eqnarray}
\label{eq:mat_SMS4_3}
\langle a |
  r^{-1} \left(\sigma_x \otimes (\sigma_{r} \boldsymbol C^{1}_q )\right) 
   | c \rangle  \nonumber \\
 =  i \int_{0}^{\infty} \frac{dr}{r}\left( Q_a P_c \langle - \kappa_ {a} m_a |
    \boldsymbol C^{1}_q | - \kappa_{c} m_c \rangle 
 -   Q_c P_a \langle \kappa_ {a} m_a |
    \boldsymbol C^{1}_q | \kappa_{c} m_c \rangle \right) \; ,
\end{eqnarray}
and the r.m.e.~\eref{eq:mat_SMS4}, 
 the third contribution to the effective interaction strength takes the form
\begin{eqnarray}
\label{eq:mat_SMS9}
X^{1}_{3} (abcd)
   =   & - &
  \langle \kappa_ {b} \|
    \boldsymbol C^{1} \| \kappa_{d} \rangle  \langle \kappa_ {a} \|
    \boldsymbol C^{1} \| \kappa_{c} \rangle
\mathpzc{V} (n_{b} \kappa_{b} , n_{d} \kappa_{d} )
 \nonumber\\
& \times &
\int_{0}^{\infty} dr  \left(\frac{- \alpha Z}{r}\right) (Q_a P_c  - Q_c P_a) \; ,
\end{eqnarray}
where we take advantage of 
$ \langle -\kappa_a \|\boldsymbol C^k \| -\kappa_c \rangle 
= \langle \kappa_a \|\boldsymbol C^k \| \kappa_c \rangle $.
\subsection{Useful one-electron reduced matrix elements}
Equations (\ref{eq:mat_NMS}), (\ref{eq:mat_SMS5}), (\ref{eq:mat_SMS8}) and (\ref{eq:mat_SMS9}) are the final key expressions of the relativistic mass shift one-electron r.m.e. that involve the following three reduced  angular one-electron matrix elements 
\begin{eqnarray}
\label{eq:d}
    \langle \kappa_a \|\boldsymbol C^1 \| \kappa_c \rangle
&=& 
    (-1)^{j_a +1/2} \; \sqrt{[j_a, j_c]}
       \left(
   \begin{array}{ccc}
      j_a   & 1 & j_c \\
      1/2 & 0 & -1/2 
   \end{array}
   \right)
      \pi \left( l_a, l_c, 1\right),
\\\nonumber\\
\label{eq:e}
    \langle -\kappa_a \|\boldsymbol \sigma^1 \| \kappa_c \rangle
&=& 
\delta(\widetilde l_a, l_c)
   (-1)^{\widetilde l_a +1/2+j_a+1} \; \sqrt{6[j_a, j_c]}
       \left\{
  \begin{array}{ccc}
      1/2  & j_a & \widetilde l_a\\
      j_c & 1/2 & 1 
   \end{array}
  \right\},
\\\nonumber\\
\label{eq:f}
    \langle \kappa_a \|\boldsymbol \sigma^1 \| -\kappa_c \rangle
&=& 
\delta( l_a, \widetilde l_c)
  (-1)^{ l_a +1/2+j_a+1} \; \sqrt{6[j_a, j_c]}
      \left\{
     \begin{array}{ccc}
   1/2  & j_a &  l_a\\
      j_c & 1/2 & 1 
  \end{array}
   \right\},
 \end{eqnarray}
where $\pi \left( l_a, l_c, 1\right)$ is defined by :
\begin{eqnarray}
\label{eq:pi}
  \pi \left( l_a, l_c, 1\right) =\left\{
  \begin{array}{ll}
  1  & \mbox{ if } l_a+1+l_c \mbox{ even,} \\ 
  0  & \mbox{ otherwise. }
  \end{array}
  \right. 
\end{eqnarray}

\section{Applications}

We wrote a new program, hereafter referred as  \textsc{rms2},  for estimating the expectation values of the relativistic nuclear recoil operators using MCDHF wave functions calculated with the \textsc{grasp2K} package~\cite{Jonetal:07a}. This code is based on the previous program \textsc{sms92}~\cite{JonFro:97a}  in which
\begin{itemize}
\item for the NMS, the one-electron radial integrals  (expression~(39)~of the original paper~\cite{JonFro:97a}) are replaced by the corresponding relativistic expression (\ref{eq:mat_NMS}), 
\item for the SMS, the first contribution $X^{1}_{1} (abcd)$ (expression~(40)~of the original paper~\cite{JonFro:97a}) is corrected by adding the relativistic contributions  (\ref{eq:mat_SMS8}) and (\ref{eq:mat_SMS9}). 
\end{itemize}

It is important to notice that the program \textsc{sms}92 calculates the uncorrected NMS as the expectation value 
$\langle\sum_i T_i \rangle$,
where $T_i$ is the Dirac kinetic energy operator  $T_i=~c~\boldsymbol\alpha_i~\cdot~\boldsymbol p_i~+~(\beta_i~-~1~)~c^2$ associated to electron~$i$, while the program \textsc{rms2} uses more accurately $\langle H^1_{\rm NMS}\rangle= \langle\sum_i p^2_i/2M \rangle$, which is consistent with section 2.1. An equivalent version has been written for the code \textsc{mcdf}-gme.

In the present work, we evaluate the NMS and SMS parameters \eref{eq:NMS} and \eref{eq:SMSred} for some low-lying levels of neutral lithium, boron-like argon and two medium-$Z$ carbon-like ions (Ca~XV and Sc~XVI) to investigate the importance of the relativistic corrections. The nuclear charge distribution is described by a Fermi model. 
 Nuclear masses  $(M_N)$ are calculated by taking away the mass of the electrons and the binding energy from the atomic mass $(M_A)$, using the formula: 
\begin{eqnarray}\label{nu_mass}
M_N(A,Z) = M_A(A,Z)-Zm_e +B_{el}(Z)
\end{eqnarray}
where the total binding energy of the electrons (expressed in eV) is estimated using~\cite{Huaetal:76a,Lunetal:2003a}
\begin{eqnarray}\label{Bind_ell}
B_{el}(Z)=14.4381~Z^{2.39} + 1.55468\cdot10^{-6}~Z^{5.35} \; .
\end{eqnarray}
The atomic and nuclear masses relevant to the present work  are reported in Tables \ref{TABLE_mass}.
\begin{table}[htdp]
\caption{\label{TABLE_mass} Atomic masses $(M_A)$~\cite{NISTAtW2} and nuclear masses $(M_N)$  (in u)  calculated from \eref{nu_mass} and \eref{Bind_ell} for lithium and argon isotopes.}
\begin{indented}
\item[]
\begin{tabular}{@{}cccll}  \br 
Isotope &\multicolumn{1}{c}{ $M_A$ } &\multicolumn{1}{c}{ $M_N$ }\\
\hline \\
$^6$Li&6.015122795(16)&6.01386737\\
$^7$Li&7.01600455(8)	&7.01474907\\
&& \\
$^{36}$Ar & 35.967545106(29)	& 35.9576862 \\
$^{40}$Ar &39.9623831225(29)	& 39.9525242\\
\br
\end{tabular} 
\end{indented}
\end{table}

\newpage
When discussing a transition mass isotope shift, one needs to consider the variation of the mass parameter from one level to another. 
The line $k$ frequency isotope shift, $\delta \nu_k^{A_1,A_2}=(\delta E_j^{A_1,A_2}-\delta E_i^{A_1,A_2})/h$, between the isotopes $A_1$ and $A_2$, of nuclear masses $M_1$ and $M_2$ respectively,  is usually written as the sum of the normal mass shift (NMS), specific mass shift (SMS) and field shift (FS) contributions :
\begin{eqnarray} \label{split_freq}
\delta \nu_k^{A_1,A_2}=\underbrace{\delta \nu_{k,{\rm NMS}}^{A_1,A_2}+\delta \nu_{k,{\rm SMS}}^{A_1,A_2}}_{{\delta \nu_{k,{\rm MS}}^{A_1,A_2}}}+\delta \nu_{k,{\rm FS}}^{A_1,A_2},
\end{eqnarray}
with 
\begin{eqnarray}\label{DeltaK}
\delta \nu^{A_1,A_2}_{k,{\rm MS}}&=&\left(\frac{M_2-M_1}{M_1M_2}\right)\frac{\Delta K_{\rm MS}}{h}
=\left(\frac{M_2-M_1}{M_1M_2}\right)\Delta \widetilde K_{\rm MS},
\end{eqnarray}
where  $\Delta K_{\rm MS}$ is the difference of the $K_{\rm MS}$ parameters of the levels involved in transition~$k$.  As far as conversion factors are concerned,
 we use\footnote{This conversion factor is calculated as $(m_e/u)2R_\infty c~\times 1.10^{-9}=3609.4824$ using the 2006 CODATA recommended values of the fundamental physical constants \cite{Mohetal:08a}.}  $\Delta \widetilde K_{\rm MS}[{\rm GHz~u}]=3609.4824~\Delta K_{\rm MS}$[$m_e E_{\rm h}$]. Note that thanks to the separability enhanced in~\eref{eq:separation},~\eref{DeltaK} can be applied to both the mass contributions NMS and SMS, separately.

\subsection{Hydrogen-like selenium}
Below we present some relevant calculations of the expression (\ref{eq:mat_NMS}) for a heavy one-electron ion (Se~XXXIV, $Z=34$). This choice is motivated by the interesting comparison with the unpublished work of Kozlov~ \cite{Koz_site}.
The normal mass shift values calculated with the operators $H^1_{\rm NMS}$ and $\left(H^2_{\rm NMS}+ H^3_{\rm NMS}\right)$, using the  \textsc{rms2} program, are reported in  Table~\ref{TABLE_I}. In the second and third column respectively, comparison is made with the numerical results of Kozlov together with our analytical values.   The latter are based on analytical hydrogenic wave functions\footnote{\label{foot}The values reported in Tables \ref{TABLE_I} and \ref{TABLE_II} are based on  $\alpha^{-1}=137.035989500$ adopted  in \textsc{grasp2K}. For $1s_{1/2}$, the analytical result for $K_{\rm NMS}$ becomes 656.358886872 if adopting the 
$\alpha^{-1}=137.035999679$ 2006~CODATA value~\cite{Mohetal:08a} .}~\cite{Joh:2007a}. The agreement is very satisfactory.

\begin{table}[htdp]
\caption{\label{TABLE_I}Contributions to the normal mass shift $K_{\rm NMS}$ parameters (in $m_e E_{\rm h}$) for  hydrogen-like selenium ($Z=34$).}\lineup
\begin{indented}
\item[]
\begin{tabular}{@{}lrrrl} \br 
&   \multicolumn{1}{c}{ \textsc{rms}2 }  & \multicolumn{1}{c}{Kozlov \cite{Koz_site}} & \multicolumn{1}{c}{Analytic} \\
\hline\\
&\multicolumn{3}{c}{$K^1_{\rm NMS}$}\\\\
$1s_{1/2}$        &   $\m656.3589797$ &$\m656.3589$&$\m656.358899684$\\
$2p_{1/2}$        & $\m154.8937900$ & $\m154.8938$ &$\m154.893789883$\\
$2p_{3/2}$        & $\m147.5192507$ & $\m147.5192$ &$\m147.519250700$\\
\hline \\
 &\multicolumn{3}{c}{$K^{2}_{\rm NMS}+K^{3}_{\rm NMS}$}\\\\
$1s_{1/2}$       &$-78.3588949$ & $-78.3589$ &$-78.3588996839$\\
$2p_{1/2}$       &$-8.0987869$ &$-8.0988$&$-8.0987868710$ \\
$2p_{3/2}$       &$-3.0192507$ & $-3.0193$ &$-3.0192506997$\\ 
\br 
\end{tabular} 
\end{indented}
\end{table}
\newpage
\subsection{Lithium-like systems using Dirac one-electron wave functions}
The SMS parameters for Li-like iron ($Z=26$) and selenium ($Z=34$) are calculated  in the single configuration approximation using three-electron wave functions built on unscreened Dirac solutions. 
The results are reported in Table \ref{TABLE_II} and compared with independent estimations using an adapted version of  \textsc{mcdf}-gme \cite{IndDes:08a,Nazetal:2011a} and with the analytical results.
The three sets are consistent with each other but sensitively different from Kozlov's values~ \cite{Koz_site} reported in the last column of the table. 
Note that the comparison is somewhat  unfair to Kozlov since the grid parameters used for the discrete representation of orbital wave functions have been adapted in both programs (\textsc{rms}2 and \textsc{mcdf}-gme)
 to achieve a better accuracy.
\begin{table}[htdp]
\caption{\label{TABLE_II}Contributions to the specific mass shift $K_{\rm SMS}$ (in $m_e E_{\rm h}$) parameters for Li-like iron ($Z=26$) and selenium~($Z=34$) using unscreened Dirac one-electron wave functions. }
\begin{indented}
\item[]
\begin{tabular}{clrrrr}  
\br
&&  \textsc{rms}2 & \textsc{mcdf}-gme&Analytic& Kozlov \cite{Koz_site}  \\
\hline\\
\multicolumn{2}{l}{Li-like iron}&\multicolumn{4}{c}{$K^1_{\rm SMS}$}
\\\\
& $1s^2 2p_{1/2}~^2P^o_{1/2}$  
    & $ -55.24725067$&$-55.24725061$&$-55.247250683$ & $-55.2474$    \\
& $1s^2 2p_{3/2}~^2P^o_{3/2}$
    & $-53.26443137$&$-53.26443136$&$-53.264431362$ & $-53.2645$\\
\\& &\multicolumn{4}{c}{$K^{2}_{\rm SMS}+K^{3}_{\rm SMS}$}\\\\
& $1s^2 2p_{1/2}~^2P^o_{1/2}$      
&$3.48269304$  &$3.48269307$ &$3.482693070$& $3.4278$ \\
& $1s^2 2p_{3/2}~^2P^o_{3/2}$    
&$1.20278262$ &$1.20278262$ &$1.202782617$& $ 1.1960$ \\
\hline
\\
\multicolumn{2}{l}{Li-like selenium }&\multicolumn{4}{c}{$K^1_{\rm SMS}$}\\\\
& $1s^2 2p_{1/2}~^2P^o_{1/2}$      
& $ -97.71464163$ &$-97.71464140$ &$-97.714641685$&   $-97.7150$ \\
& $1s^2 2p_{3/2}~^2P^o_{3/2}$      
& $-91.70637651$ &$-91.70637651$ &$-91.706376511$& $-91.7069$ \\\\
&&\multicolumn{4}{c}{$K^{2}_{\rm SMS}+K^{3}_{\rm SMS}$}\\\\
& $1s^2 2p_{1/2}~^2P^o_{1/2}$
&$10.53884731$  & 10.5388474&$10.53884746$& $10.2546$\\
&$1s^2 2p_{3/2}~^2P^o_{3/2}$
&$3.55081371$&3.5508137&$3.55081372$ & $~~3.5164$\\
\br
\end{tabular} \\
\end{indented}
\end{table}%

\newpage
\subsection{Neutral lithium in the MCDHF approach}
The MCDHF active space method consists in writing the total wavefunction as a configuration state function 
expansion built on a set of active one-electron orbitals.  To investigate the convergence of the property, the orbital set is systematically expanded up to $n = 10$, but imposing the angular restriction $l_{max}=6$
 ($i$ orbitals). The sequence of CSFs Active Spaces (AS) is resumed as follows
\begin{eqnarray}
\label{ASF_AS0}
\nonumber
\rm{AS}_{0}  & = & 1s^2 2s, \\  
\nonumber
\rm{AS}_{2} & = & {\rm AS}_{0} + \left\{ 2p \right\}, \\
\nonumber
\rm{AS}_{3} & = & {\rm AS}_{2} + \left\{ 3s, 3p, 3d  \right\}, \\
\nonumber
\rm{AS}_{4} & = & {\rm AS}_{3} + \left\{ 4s, 4p, 4d, 4f  \right\}, \\
\nonumber
\rm{AS}_{5} & = & {\rm AS}_{4} + \left\{ 5s, 5p, 5d, 5f, 5g  \right\}, \\
\nonumber
\rm{AS}_{6} & = & {\rm AS}_{5} + \left\{ 6s, 6p, 6d, 6f, 6g, 6h  \right\}, \\
\nonumber
\rm{AS}_{7} & = & {\rm AS}_{6} + \left\{ 7s, 7p, 7d, 7f, 7g, 7h, 7i  \right\}, \\
\nonumber
\rm{AS}_{8} & = & {\rm AS}_{7} + \left\{ 8s, 8p, 8d, 8f, 8g, 8h, 8i \right\}, \\
\nonumber
\rm AS_{9} & = & {\rm  AS}_{8} + \left\{ 9s, 9p, 9d, 9f, 9g, 9h, 9i \right\}, \\
\nonumber
\rm AS_{10} & = & {\rm  AS}_{9} + \left\{ 10s, 10p, 10d, 10f, 10g, 10h, 10i \right\}, 
\end{eqnarray}
where the $(nl)$-notation implies the relativistic shell structure $j=l \pm 1/2$.
The configuration space is increased progressively, by adding at each step a new layer of variational orbitals,  keeping the previous ones frozen from the $(n-1)$ calculation. The MCDHF expansions are based on single and double (SD) excitations from the configuration reference. Triple excitations are investigated through SDT-configuration interaction (CI) calculations. 
\begin{table}[htdp]
\caption{\label{Li_NMS}Uncorrected ($K^1_{\rm NMS}$) and corrected ($K_{\rm NMS}$) normal mass shift parameters  (in $m_e E_{\rm h}$) for Li I.}
\begin{indented}
\item[]
\begin{tabular}{lcccccc} \br 
\multicolumn{1}{l}{$AS_n$} 
& \multicolumn{2}{c}{$SD$} & &  \multicolumn{2}{c}{$SDT$}   \\ 
\cline{2-3} \cline{5-6}
&$ K^1_{\rm NMS}$&$K_{\rm NMS}$&& 
$ K^1_{\rm NMS}$&$K_{\rm NMS}$\\ \hline
\multicolumn{4}{c}{$1s^22s~^2S_{1/2}$} \\
$n$=5  & 7.479955285& 7.473188966&& 7.480387512& 7.473620714\\
$n$=6  & 7.480757179& 7.473989593&& 7.481294538& 7.474526401\\
$n$=7  & 7.480843823& 7.474076167&& 7.481413156& 7.474644913\\ 
$n$=8  & 7.482617678& 7.475849092&& 7.483709525& 7.476940080\\ 
$n$=9  & 7.482764972& 7.475996085&& 7.483865298& 7.477095534\\ 
$n$=10 & 7.482767804& 7.475998981&& 7.483872626& 7.477102933\\ 
\\
\multicolumn{4}{c}{$1s^22p~^2P^o_{1/2}$} \\
$n$=5  & 7.411878601& 7.405201316&& 7.412125687& 7.405448151\\
$n$=6  & 7.412307495& 7.405629843&& 7.412599631& 7.405921698\\
$n$=7  & 7.412593434& 7.405916172&& 7.413034981& 7.406357367\\ 
$n$=8  & 7.414193990& 7.407516244&& 7.415203718& 7.408525555\\ 
$n$=9  & 7.414351543& 7.407673608&& 7.415377292& 7.408698936\\ 
$n$=10 & 7.414365017& 7.407687009 &&7.415402512& 7.408724081    \\ 
\\
\multicolumn{4}{c}{$1s^22p~^2P^o_{3/2}$} \\
$n$=5  & 7.411871260& 7.405208436&& 7.412118064& 7.405455006\\
$n$=6  & 7.412300503& 7.405637317&& 7.412592555& 7.405929108\\
$n$=7  & 7.412584979& 7.405922413&& 7.413026010& 7.406363146\\ 
$n$=8  & 7.414185599& 7.407522728&& 7.415193271& 7.408530126\\ 
$n$=9  & 7.414343399& 7.407680348&& 7.415366987& 7.408703663\\ 
$n$=10 &7.414356793& 7.407693666 &&7.415392260& 7.408728857   \\ 
\br 
\end{tabular}
\end{indented}
\end{table}

\begin{table}[htdp]
\caption{\label{Li_SMS}Uncorrected ($K^1_{\rm SMS}$) and corrected ($K_{\rm SMS}$) specific mass shift parameters  (in $m_e E_{\rm h}$) for Li I.} 
\begin{indented}
\item[]
\begin{tabular}{lccccc} \br 
\multicolumn{1}{l}{$AS_n$} & \multicolumn{2}{c}{$SD$} & &  \multicolumn{2}{c}{$SDT$}  \\ \cline{2-3} \cline{5-6}
&$ K^1_{\rm SMS}$&$ K_{\rm SMS}$&&$ K^1_{\rm SMS}$&$ K_{\rm SMS}$\\ \hline
\multicolumn{3}{c}{$1s^22s~^2S_{1/2}$} \\
$n$=5  & 0.3010343291 & 0.3008225633 && 0.3013767853 & 0.3011648528\\
$n$=6  & 0.3010361585 & 0.3008243666 && 0.3014579841 & 0.3012459847\\
$n$=7  & 0.3019544951 & 0.3017423943 && 0.3024396569 & 0.3022273237\\ 
$n$=8  & 0.3018617523 & 0.3016497153 && 0.3024115843 & 0.3021992791\\ 
$n$=9  & 0.3017987398 & 0.3015867742 && 0.3023512554 & 0.3021390203\\ 
$n$=10 & 0.3018561821 & 0.3016442119 && 0.3024141615 & 0.3022019200\\ 
\\
\multicolumn{3}{c}{$1s^22p~^2P^o_{1/2}$} \\
$n$=5  & 0.2489342617 & 0.2487564343 && 0.2490604867& 0.2488826064\\
$n$=6  & 0.2482881614 & 0.2481107089 && 0.2484282378& 0.2482507247\\
$n$=7  & 0.2482993836 & 0.2481222326 && 0.2484107486& 0.2482335819\\ 
$n$=8  & 0.2476015693 & 0.2474248242 && 0.2474557225& 0.2472791143\\ 
$n$=9  & 0.2475207029 & 0.2473440589 && 0.2473719731& 0.2471954806\\ 
$n$=10 & 0.2476566450 & 0.2474799659 &&0.2475131224& 0.2473365865  \\ 
\\\multicolumn{3}{c}{$1s^22p~^2P^o_{3/2}$} \\
$n$=5  & 0.2489331884 & 0.2487377216 && 0.2490592224& 0.2488636586\\
$n$=6  & 0.2482892038 & 0.2480941432 && 0.2484289962& 0.2482338261\\
$n$=7  & 0.2483022860 & 0.2481070388 && 0.2484139538& 0.2482185782\\ 
$n$=8  & 0.2476039924 & 0.2474089620 && 0.2474585594& 0.2472634414\\ 
$n$=9  & 0.2475233735 & 0.2473284034 && 0.2473750253& 0.2471799634\\ 
$n$=10 & 0.2476606363 & 0.2474656360 &&0.2475176569& 0.2473225644\\ 
\br 
\end{tabular}
\end{indented}
\end{table}
Tables \ref{Li_NMS} and \ref{Li_SMS} present the evolution of the NMS and the SMS parameter, respectively.
In each table both the uncorrected ($K^1_{\rm MS}$) and corrected ($K_{\rm MS}$) values are reported. 
Comparing the SD and SDT calculations, we observe that the influence of the triple excitations reaches more than 1 \% for the SMS while it is one order of magnitude smaller (0.1\%) for the NMS. 

In Table \ref{tab:kozh_comp} the individual contributions to the mass shift $\Delta\widetilde K_{\rm MS}~(=\Delta K_{\rm MS}/h$) para\-meters 
are reported for the $2p_{1/2}~^2P^o_{1/2}-2s~^2S_{1/2}$ ~($D_1$~line) and  $2p_{3/2}~^2P^o_{3/2}-2s~^2S_{1/2}$~($D_2$~line)  transitions in lithium. 
Values are calculated with the SD and SDT $n=10$ active space final  results of Tables \ref{Li_NMS} and~\ref{Li_SMS}. 
Although  many robust theoretical studies on the resonance line transition isotope shifts are available (see Table~\ref{tab:kozh_comp_tot} and discussion below), the comparison with other theoretical works presented  in Table \ref{tab:kozh_comp}  is limited to the recent large-scale configuration-interaction Dirac-Fock-Sturm calculations of Kozhedub \etal \cite{Kozetal:2010a} since these authors precisely focus on the estimation of the relativistic nuclear recoil corrections. Kozhedub \etal 's values are very consistent with our results: They report $\Delta(\widetilde K^2_{\rm NMS}+\widetilde K^3_{\rm NMS}) = 0.33$ and 0.38~GHz~u for the $D_1$ and $D_2$ transitions respectively. However, the uncorrected NMS contribution and  therefore, the total NMS values, sensitively differ from each other by around 1.6~GHz~u. This latter discrepancy is not understood yet and clearly deserves further investigations. 

 The uncorrected contribution of the SMS is also compared with the non-relativistic result of Godefroid~\textit{et al}~\cite{Godetal:01a} using the multiconfiguration Hartree-Fock method. More interesting is the comparison with the recent SMS values of Kozhedub \etal \cite{Kozetal:2010a} investigating the relativistic recoil corrections and using the same NMS and SMS partition according to~(\ref{eq:H_NMS}) and (\ref{eq:H_SMS}). As for the NMS, the relativistic corrections are in very nice agreement (they report 
 $\Delta (\widetilde K^2_{\rm SMS}+\widetilde K^3_{\rm SMS}) =  0.12$ and 0.06~GHz~u for the $D_1$ and $D_2$ lines ) but the uncorrected forms do differ substantially with our estimation  
 (they report for instance for the $D_1$ line, $\Delta\widetilde K^1_{\rm SMS} = - 198.78$, against our value of $ -198.164$~GHz~u).


\clearpage
The comparison with observation for the individual mass contributions is also limited. There are a few reasons for this. First, as illustrated by \eref{split_freq}, the field shift contribution should be properly subtracted from the observed transition frequency before trying to extract the mass contribution. But this is usually the other way round that makes the theoretical calculation of mass shifts  interesting: for a few-electron atomic systems like lithium indeed, the difference between the mass contribution calculated by elaborate {\it ab initio} calculations and the observed transition IS allows to extract the change in the mean square charge radius of the nuclear charge distributions for all  isotopes, as illustrated by the very recent and complete work of N\"ortersh\"auser \etal~\cite{Noretal:2011a}. Another good reason is that once the FS ``eliminated'', a clean separation of the NMS and SMS contributions could be criticized, as pointed out by Palmer~\cite{Pal:87a}. However,  remembering that for lithium, the FS is roughly 10$^4$ times smaller than the MS,  it is worthwhile to neglect it for trying the mass separation exercise.
There is indeed one experimental work by Radziemski \etal~\cite{Radetal:95a} discussing the NMS and SMS separation in this line but as we will observe later (see Table~\ref{tab:kozh_comp_tot}), the corresponding experimental transition IS values are not aligned with most of the other observed values. 
In their work, these authors separate the two mass shift  contributions from the experimental transition IS in $^{6,7}$Li, neglecting the field shift contribution and approximating the Bohr mass shift by the experimental observed level energy, as suggested by M\aa rtensson and Salomonson \cite{MarSal:82a}, 
\begin{eqnarray}\Delta^{\rm BMS}=-\frac{m_e}M E_M^B\simeq-\frac{m_e}M E_{exp}\;.
\end{eqnarray}
From the same expression, we  build the transition Bohr mass shift for the $^{6,7}$Li isotope pair
\begin{eqnarray}\label{eq:NMS_expe}
\delta E_{\rm BMS}\simeq\left(\frac{m_e}{M(^6Li)}-\frac{m_e}{M(^7Li)}\right)\Delta E_{exp}\;
\end{eqnarray}
from the obserbed transition  energy. Combining~\eref{DeltaK} and \eref{eq:NMS_expe}, one finds
\begin{eqnarray}\label{BMS}
\Delta\widetilde K_{\rm NMS}
\simeq m_e~\frac{\Delta E_{exp}}{h} 
= m_e \; \nu_{exp}
\end{eqnarray}
from which we estimate the ``observed'' NMS values reported in Table \ref{tab:kozh_comp}, using the most recent absolute frequency measurements of Das and Natarajan \cite{Dasetal:2007a}.
The corresponding ``observed''  $\Delta \widetilde K_{\rm SMS}$ values are calculated by substracting the so-estimated NMS contribution from the experimental IS line shifts 
($-443.9490(16)$~GHz~u and $-443.9126$(29)~GHz~u, for $D_1$ and D$_2$, respectively).
Note that we did not take the liberty of reporting the frequency uncertainties estimated by Das and Natarajan on the separate contributions, the separability of NMS and SMS being by itself questionable.
\begin{table}[htdp]
\caption{\label{tab:kozh_comp}Individual contributions to the mass shift $\Delta\widetilde K_{\rm MS}$ (GHz u) parameters for the $2p~^2P^o_{1/2}~-~2s~^2S_{1/2}$ and  $2p~^2P^o_{3/2}-2s~^2S_{1/2}$ transitions in lithium}\lineup
\vspace{1mm}
\begin{indented}
\item[]
\begin{tabular}{lllrr} \br \\
&&&\multicolumn{1}{c}{$~^2P^o_{1/2}~-~^2S_{1/2}$} &  
\multicolumn{1}{c}{$~^2P^o_{3/2}-~^2S_{1/2}$}\\
\\\hline\\
NMS&$\Delta\widetilde K^1_{\rm NMS}$ 	&SD	&$-246.899$&	$-246.928$\\
&	&SDT&$-247.142$&	$-247.179$\\
&$\Delta(\widetilde K^2_{\rm NMS}+\widetilde K^3_{\rm NMS})$
&	SD	&$0.328$&	$0.382$\\
 &        &SDT &$0.333$&	$0.384$\\
\\&$\Delta\widetilde K_{\rm NMS}$ &SDT&$-246.812$&$-246.795$\\       
&\\
&Other theory$^a$&        	&$-245.15\0$		&$-245.11\0$	\\\\
&Obs.$^b$&&$-245.103$&$-245.108$\\  
\\\hline\\
SMS&$\Delta\widetilde K^1_{\rm SMS}$          
	&SD	&$-195.632$	&$-195.618$\\
         &&SDT	&$-198.164$	&$-198.148$\\
&$\Delta (\widetilde K^2_{\rm SMS}+\widetilde K^3_{\rm SMS})$
&SD		&$0.127$&	$0.061$\\
&&SDT		&$0.129$&	$0.062$\\
\\&$\Delta \widetilde K_{\rm SMS}$          
&	SDT	&$-198.035$	&$-198.086$\\\\
&Other theory$^a$&        	&$-198.78\0$		&$-198.77\0$	\\
&Other theory  (NR)$^c$&        	&$-198.66\0$		&$-198.71\0$	\\
\\
&Obs.$^d$ 
&	&$-198.843$	&$-198.101$\\\\
\br
\end{tabular}
\\$~^a$CI Dirac-Fock-Sturm calculation of Kozhedub \etal \cite{Kozetal:2010a}.
\\$~^b$NMS values deduced from the transition frequencies \cite{Dasetal:2007a} using \eref{BMS} (see text).\\
$~^c$Non-relativistic MCHF calculations \cite{Godetal:01a}.\\
$~^d$SMS values obtained by subtracting the ``observed'' NMS (see footnote $b$ above) from the~IS measured by Das and Natarajan \cite{Dasetal:2007a}.
\end{indented}
\end{table}

Cleaner and in principle less problematic should be the comparison of the total mass shifts, as reported in Table \ref{tab:kozh_comp_tot}.  On the theoretical side, we refer to the study of Korol and Kozlov~\cite{KorKoz:07a} 
treating electron correlation with configuration interaction (CI) and many-body perturbation theory (MBPT) methods with Dirac-Fock orbitals, 
to the calculations of Kozhedub \etal\cite{Kozetal:2010a} using large-scale configuration-interaction Dirac-Fock-Sturm method and to the  Yan \etal\cite{Yanetal:08a} calculations estimating the mass corrections from highly correlated non-relativistic wave functions expressed in Hylleraas coordinates\footnote{The values of Yan \etal reported in the Kozhedub's paper \cite{Kozetal:2010a} suggest that the atomic mass has been used in order to evaluate the mass shift parameter. In table \ref{tab:kozh_comp_tot}, the Yan \etal's values have been reevaluated using the nuclear mass.}. 
From all these elaborate results, we only kept the mass contributions, systematically excluding the contributions from the nuclear size corrections.
We already noticed the differences between Kozhedub \etal's results and ours appearing in the separate NMS and SMS contributions. As commented above, these differences do not arise from the relativistic corrections ($K^2 + K^3$), but rather from the ``uncorrected''$K^1$ values,  and should be further investigated.
Our results seem to be of higher quality than the CI+MBPT results of Korol and Kozlov. As far as the differences with Yan \etal's results are concerned, we should keep in mind i) that our orbital active set is truncated to $l_{max}= 6$, ii) that the layer approach adopted in the SD-MCDHF optimization could be a limiting factor and  iii) that the convergence of the  $\Delta\widetilde K_{\rm MS}$ parameter as a function of the size of the active set is slow and not yet achieved at $n=10$, as illustrated by the comparison of the  two $n=9$ and $n=10$ sets of  results reported in the Table~\ref{tab:kozh_comp_tot}.


On the experimental side,  we display in the same Table~\ref{tab:kozh_comp_tot},  the experimental isotope shift values somewhat abusively converted in $\Delta\widetilde K_{\rm MS}$ parameters, ie. neglecting the FS contribution and inverting~\eref{DeltaK},  $\Delta \widetilde K_{\rm MS} = \delta \nu_{k}(M_1M_2)/ (M_2-M_1)$.  As already mentioned, this conversion  is unfair to physicists who do some huge efforts to extract the nuclear charge radii from the FS~\cite{Noretal:2011a},  but has the merit of  illustrating where the present  modest contribution  lies in the distribution of experimental values.  
From this not exhaustive chronological list (\cite{Sanetal:95a,Radetal:95a,Schetal:96a,Busetal:2003a,Waletal:2003a, Nobetal:2006a, Dasetal:2007a}), it is clear that Radziemski \etal's results  lie a bit outside the experimental distribution. 

\begin{table}[htdp]
\begin{indented}
\item[]
\caption{\label{tab:kozh_comp_tot}Mass shift $\Delta\widetilde K_{\rm MS}$ (GHz u) for the $2p_{1/2}~^2P^o_{1/2}-2s~^2S_{1/2}$ and  $2p_{3/2}~^2P^o_{3/2}-2s~^2S_{1/2}$ transitions in lithium, compared with experimental IS.} 
\begin{tabular}{lllll} \br \\
&\multicolumn{1}{c}{$~^2P^o_{1/2}~-~^2S_{1/2}$} &  
\multicolumn{1}{c}{$~^2P^o_{3/2}-~^2S_{1/2}$}&\multicolumn{1}{c}{Ref.}\\
\\  \hline \\
$n=9$     &$-445.1941$   &$-445.2330$&This work\\ 
$n=10$ &$-444.8442$  &$-444.8808$&This work\\ 
 
\\
Other theory	&$-447(12)$		&$-447(12)$		&Korol and Kozlov \cite{KorKoz:07a}\\
			&$-443.81(20)$		&$-443.82(20)$		&Kozhedub \etal\cite{Kozetal:2010a}\\
			&$-443.860337(253)$	&$-443.876984(253)$	&Yan \etal\cite{Yanetal:08a}\\
\\
Experiment$^a$	
			&$-443.89(3)$		&$-443.91(2)$		&Sansonetti  \etal\cite{Sanetal:95a}\\
			&$-443.46(63)$	&$-443.59(63)$	&Radziemski \etal \cite{Radetal:95a}\\
			&$-443.9033(63)$	&$-443.9791(63)$	&Scherf \etal\cite{Schetal:96a}\\
			&$-443.9045(29)$	&				&Bushaw\etal\cite{Busetal:2003a}\\
			&$-443.951(5)$	&				&Walls  \etal\cite{Waletal:2003a}\\
			&$-443.941(3)$	&$-443.948(4)$	&Noble \etal \cite{Nobetal:2006a}\\
			&$-443.9490(16)$	&$-443.9126(29)$	&Das and Natarajan \cite{Dasetal:2007a}\\
\br
\end{tabular}
\\$~^a$ inverting~\eref{DeltaK}, ie. using
 $\Delta \widetilde K_{\rm MS} = \delta \nu (M_1M_2)/ (M_2-M_1) $ (see text)
\end{indented}
\end{table}
\subsection{B-like argon}

Large-scale calculations are performed for $1s^22s^2 2p ~^2P^o_{1/2,3/2}$ of B-like argon ($Z=18$).
The radial orbital basis is obtained from SD-MCDHF calculations, including single and double excitations from all shells of the $\{1s^22s^2 2p,1s^2 2p^3\}$ complex to increasing orbital active sets, up to the 
\{$10s9p8d7f6g3h1i$\}.
Subsequently to this  layer-by-layer SD-MCDHF orbital optimization, RCI calculations are performed including the Breit and QED effects in a space generated by SD excitations from the extended \{$1s^22s^22p$, $1s^22p^3$, $1s^22s2p3d$, $1s^22p3d^2$\} multireference set to the full orbital set. The expansion for the two~$J$~values includes more than 200~000 relativistic CSFs. This computational strategy has been developed by Rynkun \etal \cite{Rynetal:2011a} for the evaluation of transition rates in boron-like ions, from N~III to Zn~XXVI.

Table \ref{tab:Ar} illustrates the convergence of the NMS and SMS contributions with the increasing of the active set. 
\begin{table}[htdp]
\caption{\label{tab:Ar}NMS and SMS parameters (in $m_e E_{\rm h}$) values for the states  $1s^22s^2 2p~^2P^o_{1/2}$ and
$1s^22s^2~2p~^2P^o_{3/2}$ states of B-like Ar.}
\begin{indented}
\item[]
\begin{tabular}{lcccccc} \hline 
\\\multicolumn{1}{l}{$AS_n$} 
        & $K^1_{\rm NMS}$   &$\left(K^2_{\rm NMS}+ K^3_{\rm NMS}\right)$ &&  $ K^1_{\rm SMS}$   & $\left(K^2_{\rm SMS}+ K^3_{\rm SMS}\right)$\\
\\\hline\\
\multicolumn{5}{c}{$1s^22s^2 2p ~^2P^o_{1/2}$} \\
&&&&& \\
$n=3$&$417.6959959$&$-12.90905322$&&	$-16.14960237$&$0.46905612$\\
$n=4$&$418.1118745$&$-12.91188239$&&	$-16.30895958$&$0.47763419$\\
$n=5$&$418.2177551$&$-12.91178616$&&	$-16.33759742$&$0.47889811$\\
$n=6$&$418.2537558$&$-12.91247514$&&	$-16.32403619$&$0.47580853$\\
$n=7$&$418.2919516$&$-12.91360934$&&	$-16.34490331$&$0.47658869$\\
$n=8$&$418.2945374$&$-12.91354179$&&	$-16.34259843$&$0.47612308$\\
$n=9$&$418.2974119$&$-12.91373611$&&	$-16.34355766$&$0.47621031$\\
$n=10$&$418.2982181$&$-12.91375324$&&$-16.34352453$&$0.47609622$\\
&&&&& \\
$n=10_{expand}$&$418.2995162$&$-12.91378955$&&$-16.33796596$&$0.47600086$\\
&&&&& \\
\hline\\
\multicolumn{5}{c}{$1s^22s^2 2p ~^2P^o_{3/2}$} \\
&&&&& \\
$n=3$&$417.3566053$&$-12.66720414$&&	$-15.91469860$&$0.12821557$\\
$n=4$&$417.7946482$&$-12.66774060$&&	$-16.09016993$&$0.13464620$\\
$n=5$&$417.9013634$&$-12.66662806$&&	$-16.11510903$&$0.13335729$\\
$n=6$&$417.9378902$&$-12.66759777$&&	$-16.10142189$&$0.13128456$\\
$n=7$&$417.9773669$&$-12.66844094$&&	$-16.12255333$&$0.13135116$\\
$n=8$&$417.9799742$&$-12.66842909$&&	$-16.12002982$&$0.13104864$\\
$n=9$&$417.9828424$&$-12.66858843$&&	$-16.12097800$&$0.13104262$\\
$n=10$ & $417.9836553$  &$-12.66862952$ & & $-16.12094341$ & $0.13100231$ \\
&&&&& \\
$n=10_{expand}$&$417.9846675$&$-12.66865788$&&$ -16.11555533$&$0.13089443$\\
&&&&& \\
\hline 
\end{tabular}
\end{indented}
\end{table}
In Table \ref{tab:Tup_comp}, the isotope shifts of the forbidden transitions $1s^22s^2 2p~^2P^o_{1/2}~-~^2P^o_{3/2}$ in $^{36,40}$Ar are presented and compared with the results of Tupitsyn \etal \cite{Tupetal:03a}. In their work, the  CI Dirac-Fock method was used to solve the Dirac-Coulomb-Breit equation and to calculate the energies and the isotope shifts. The CSFs expansions were generated including ``all single and double excitations and some part of triple excitations''. The nuclear charge distribution is described by a Fermi model and is therefore consistent with the present work. The large cancellation of the terms involved in the transition isotope shift makes accurate calculations very challenging. 

Table~\ref{tab:Tup_comp} shows the individual contributions of operators~\eref{eq:H_NMS} and~\eref{eq:H_SMS} to the wavenumber mass shift. 
A good agreement is observed between the two sets of values, the total wavenumber mass shift values differing by less than 0.8\%. This example beautifully confirms the importance of the relativistic corrections to the recoil operator: the total wavenumber mass shift would be indeed 50\% 
smaller if estimated from the uncorrected form of the mass Hamiltonian $\langle H^1_{\rm NMS} + H^1_{\rm SMS} \rangle$. The nice agreement with Tupitsyn \etal's results is also a good sign of reliability for the tensorial form derivation of the nuclear recoil Hamiltonian of section~2. 
\begin{table}[htdp]
\caption{\label{tab:Tup_comp}Individual contributions to the wavenumber mass shift $\delta \sigma$ (cm$^{-1}$) for the forbidden transition $1s^2 2s^2~2p~^2 P^o_{1/2}-~^2P^o_{3/2}$ in boron-like $^{36,40}$Ar.} 
\begin{indented}
\item[]
\begin{tabular}{lcccccc} 
\br
&\multicolumn{4}{c}{$\delta \sigma$}\\
\cline{2-5}
&\multicolumn{1}{c}{$\langle H^1_{\rm NMS}\rangle$}   
&\multicolumn{1}{c}{$\langle H^2_{\rm NMS}+ H^3_{\rm NMS}\rangle$} 
&\multicolumn{1}{c}{$\langle H^1_{\rm SMS}\rangle$}   
&\multicolumn{1}{c}{$\langle H^2_{\rm SMS}+ H^3_{\rm SMS}\rangle$}&Total\\
\hline 
&&&&& \\
This work		&$0.1054$	&$-0.0821$	&$-0.0745$	&$0.1155$	& 0.0645\\
Tupitsyn \etal\cite{Tupetal:03a}	&$0.1053$		&$-0.0822$	&$-0.0742$	&$0.1151$	&0.0640\\
&&&&& \\
\br
\end{tabular}
\end{indented}
\end{table}

\subsection{C-like ions calculations}

As another illustration of the importance of the relativistic corrections to the recoil operator, the values of the SMS, NMS and total level mass shift parameters are reported in Table~\ref{Ca_Sc} for the levels arising from the ground configuration $1s^2 2s^2 2p^2$ in Ca~XV and Sc~XVI.
As far as the calculations are concerned, the orbitals are obtained by SD-MCDHF calculations, considering single and double excitations from all shells of the $\{ 1s^2 2s^22p^2, 1s^2 2p^4\}$ Layzer complex to the \{$8s7p6d5f4g2h$\} active set. These MCDHF calculations are followed by relativistic configuration interaction (RCI)  calculations, including the Breit interaction and the QED corrections, using the enlarged multireference  $\{1s^22s^22p^2, 1s^22p^4, 1s^22s2p^23d, 1s^22s^23d^2\}$ set. The size of the expansions is around 350 000 relativistic CSFs. This computational method has been used by J\"onsson \etal~\cite{Jonetal:2011a} to calculate transition rates, hyperfine structures and Land\'e $g$ factors for all carbon-like ions between F IV and Ni XXIII.

On the absolute scale of level shift parameters, one  observes that the relativistic corrections  $(K^2_{\rm NMS}+K^3_{\rm NMS})$  to the NMS  have the same order of magnitude than the uncorrected SMS contribution.
Transition isotope shifts are more interesting properties since they are the real observables if the resolution is good enough. These are monitored by the differential effects on the level~IS.
It is interesting to infer from Table~\ref{Ca_Sc}  the possible mass isotope shifts on the intraconfiguration (M1/E2) transition frequencies. 
Considering for example the Ca~XV $\; ^3P_1 \rightarrow \; ^3P_2$ transition, the uncorrected total mass shift change is enlarged by a factor two  when including the $(K^2_{\rm MS}+K^3_{\rm MS})$ relativistic corrections.
For  the $\; ^3P_1 \rightarrow \; ^1D_2$ transition, a similar increase of the mass shift is predicted but of ``only'' 20\%.
Some reduction could occur: this is the case of $\; ^3P_2 \rightarrow \; ^1S_0$ (13\%).
%
%
For the  $\; ^3P_0 \rightarrow \; ^1D_2$ transition, the  relativistic recoil corrections reach 48\%.
%

\begin{table}[htdp]
\caption{\label{Ca_Sc}Specific mass shift $K_{\rm SMS}$, normal mass shift $K_{\rm NMS}$ , total mass shifts $K_{\rm MS} $ parameters (all in $m_e E_{\rm h}$) for $2s^2 2p^2$ levels of  Ca~XV and Sc~XVI from multireference RCI calculations. $K = K^1 + K^2 + K^3$.}
\begin{indented}
\item[]
\begin{tabular}{lcccccccccc} \br
           &         & \multicolumn{3}{c}{SMS} & \multicolumn{3}{c}{NMS} & \multicolumn{3}{c}{Total} \\
Level & $J$ & $ K^1_{\rm SMS}$    &$ K_{\rm SMS}$ && $ K^1_{\rm NMS}$     &$ K_{\rm NMS}$    &&$K^1_{\rm MS}$& $K_{\rm MS}$ &$K^2_{\rm MS}+K^3_{\rm MS}$
\\ \hline\\
\multicolumn{11}{c}{Ca XV}\\
&&&&&&&&&& \\
$2s^22p^2~^3P$  & 0 & $-$41.993 & $-$40.659 && 558.358 &538.033            &&516.364        &497.374& $-$18.990 \\                  
                & 1 & $-$41.736 & $-$40.741 && 558.070 &537.983            &&516.334        &497.242& $-$19.092 \\                           & 2 & $-$41.588 & $-$40.821 && 557.854 &537.929            &&516.266        &497.108& $-$19.158 \\
$2s^22p^2~^1D$        & 2 & $-$41.451 & $-$40.756 && 557.479 &537.607            &&516.028        &496.851& $-$19.177 \\
$2s^22p^2~^1S$        & 0 & $-$41.862 & $-$41.199 && 557.000 &537.200            &&515.138        &496.001& $-$19.137 \\
&&&&&&&&&& \\
\hline\\
\multicolumn{11}{c}{Sc XVI}\\
&&&&&&&&&& \\
$2s^22p^2~^3P$  & 0 & $-$47.460 & $-$45.771 && 621.221 &596.303            &&573.761        &550.532& $-$23.229 \\
                          & 1 & $-$47.117 & $-$45.874 && 620.842 &596.236            &&573.725        &550.362& $-$23.363 \\
                          & 2 & $-$46.950 & $-$45.964 && 620.592 &596.169            &&573.642        &549.219& $-$24.423
\\
$2s^22p^2~^1D$        & 2 & $-$46.765 & $-$45.922 && 620.145 &595.826            &&573.380        &549.904& $-$23.476
\\
$2s^22p^2~^1S$        & 0 & $-$47.229 & $-$46.421 && 619.624 &595.390            &&572.395        &548.969& $-$23.426
\\
\br
\end{tabular}
\end{indented}
\end{table}
\section{Conclusion and outlook}
The irreducible tensorial form of the nuclear recoil Hamiltonian is derived  in the present work, opening interesting perspectives for calculating isotope shifts in the multiconfiguration Dirac-Hartree-Fock framework. We implemented the formalism in the relativistic package  \textsc{grasp2K} by writing a dedicated code (\textsc{rms2})  for estimating the expectation values of the relativistic nuclear recoil operators. The comparison with other works is satisfactory and the results are promising, although not achieving the accuracy of the state-of-the-art methodology available for a few-electron systems. Electron correlation remains the major problem that might be solved in our schema with  the use of ``localized pair-correlation functions interaction method", as proposed by Verdebout \etal \cite{Veretal:10a}. The present work  enhances the fact that the relativistic corrections to the nuclear recoil are definitively necessary for getting reliable isotope shift calculations.  
The new  computational tool, that we developed on the basis of the irreducible tensorial operator techniques, will hopefully provide valuable mass isotope shift data for large systems for which there are no reliable theoretical or experimental values. 

\section*{Acknowledgements}
\noindent
C\'edric Naz\'e is grateful to the ``Fonds pour la formation \`a la Recherche dans l'Industrie et dans l'Agriculture" of Belgium for a Ph.D. grant (Boursier F.R.S.-FNRS).
Michel Godefroid and C\'edric Naz\'e thank the Communaut\'e fran{\c c}aise of Belgium (Action de Recherche Concert\'ee) for financial support. MRG also acknowledges the Belgian National Fund for Scientific Research (FRFC/IISN Convention). Finally, the authors thank Jiguang Li for fruitful discussions.

\clearpage

%
%

\bibliographystyle{unsrt}
\section*{References}

%
%
\bibliographystyle{unsrt}

\end{document}